\documentclass[twocolumn,aps,pre,superscriptaddress]{revtex4-1}

\usepackage{graphicx}
\usepackage{amsfonts, amsmath, amstext, amssymb, amsfonts, amsxtra}
\usepackage{braket}
\usepackage{bbm}
\usepackage{bbold}
\usepackage[protrusion=true,expansion=true]{microtype}

\DeclareMathOperator{\tr}{tr}

\newcommand{\im}{{\rm i}}

\newcommand{\Di}{\mathcal{D}}

\newcommand{\rhop}{\hat{\rho}} 
\newcommand{\Hop}{\hat{H}} 
\newcommand{\Uop}{\hat{U}}

\newcommand{\cop}{\hat{c}^{}} 
\newcommand{\cdop}{\hat{c}^{\dagger}} 
\newcommand{\nop}{\hat{n}} 
\newcommand{\cur}{\hat{\mathcal{J}}} 
\newcommand{\cura}{\mathcal{J}}

\newcommand{\sutd}{Engineering Product Development Pillar, Singapore University of Technology and Design, 8 Somapah Road, 487372 Singapore} 
\newcommand{\majulab}{MajuLab, CNRS-UNS-NUS-NTU International Joint Research Unit, UMI 3654, Singapore}

\begin{document}

\title{Oscillation and decay of particle current due to a quench and dephasing in an interacting fermionic system}
\author{Kenny Choo} 
\affiliation{\sutd} 
\author{Ulf Bissbort} 
\affiliation{\sutd}  
\author{Dario~Poletti} 
\affiliation{\sutd} 
\affiliation{\majulab}
\date{\today}
\pacs{03.75.Gg, 03.65.Yz, 42.50.Dv}

\begin{abstract}
We study the response of a particle current to dissipative dephasing in an interacting, few-body fermionic lattice system. The particles are prepared in the ground state in presence of an artificial magnetic gauge field, which is subsequently quenched to zero. The initial current decays non-trivially in the dissipative environment and we explore the emerging dynamics and its dependence on various system parameters. 
\end{abstract}

\maketitle

\section{Introduction}\label{sec:intro}     
Many body interacting quantum systems have drawn an increasing amount of attention in recent years. Such systems can feature exotic, out-of-equilibrium phases of matter in their steady state \cite{Mitra, Zoller,Giamarchi,Cirac, GuoPoletti2016}. Moreover, even at intermediate times, interesting correlations can be generated \cite{BernierKollath}, and their relaxation dynamics towards the steady state can occur in various ways, from a simple exponential, stretched exponentials to power-law and more \cite{PolettiKollath2012, CaiBarthel2012, PolettiKollath2013, CarmeleDalmonte, SciollaKollath2015, MedvedyevaZnidaric, Garrahan, MarcuzziLesanovsky2014, MarcuzziLesanovsky2015}. A growing number of experimental groups in the ultracold atoms community has started investigating interacting quantum systems in contact with an environment \cite{Ott, Ott2, Vengalattore, Broida}.

Many body open quantum system also have interesting applications in the field of quantum thermodynamics. Specifically, understanding how many body properties of a quantum system can affect the ability to convert heat into work is one important open question, which deserves further investigation. Recent works have started exploring how the particle statistics or the strength of interactions \cite{ZhengPoletti2015, Talkner, AnotherDelCampo, Deffner} affect the properties of a thermodynamic cycle.

Another important avenue for the application of many body open quantum systems is in the context of atomtronics, that is the study of electronics-inspired systems consisting of neutral atoms \cite{Seaman, DanaAnderson}. From this line of research, the study of atoms in a ring-structured circuit is of particular importance \cite{Campbell, Campbell2}. Another key advance has been the ability to generate artificial magnetic gauge fields in (neutral) ultracold atomic systems, thus giving the opportunity to emulate the effects of a magnetic field on charged particles \cite{Spielmanandtworeviews, GaugeReview2}. 

It is in this context that we set our work. We consider a few-body system of interacting, spinless fermionic particles in a ring which are prepared in the non-dissipative ground state of the system with an artificial gauge field, characterized by a complex tunneling phase $\phi$. This initial interacting state features a current and we study its dynamics after instantaneously quenching the gauge field to zero and exposing the system to a particular type of dissipation. The dissipation chosen is able to suppress the current in the system. We will analyze a small ring (1D lattice with periodic boundary conditions) consisting of up to $8$ sites, for different fillings, number of atoms and interaction strength. We will also consider different initial currents imposed by the gauge field.      
We will show that the response of the initial current to the gauge field features jumps at specific values. The interaction induces a change from a purely exponential decay to an oscillating, exponentially suppressed decay. The frequency of the oscillations increases as the interaction becomes larger, with corrections due to the presence of dissipation. We also show that the system is much more sensitive to the strength of the interaction near half-filling, although less to the initial artificial gauge field. To gain insight, we analyze the overlap of the initial state and the eigenstates of the Hamiltonian without the artificial magnetic field.

The paper is structured as follows: In section \ref{sec:model} we introduce our model in detail, as well as our numerical implementation. In section \ref{sec:deca} we discuss the main results of the work, i.e. the dependence of the system's dynamics on the filling, the strength of interaction, the initial gauge field and of the dephasing rate. Last, in section \ref{sec:conclusions} we draw our conclusions.

\section{Model}\label{sec:model}   
We consider a dissipative system of single band spinless fermions in a ring with nearest neighbor interactions. The Hamiltonian of the system is given by
\begin{equation}\label{eq:hamiltonian}
\Hop(\phi) = -J\sum_{\langle l,l' \rangle} \left( e^{-\im \phi}\hat{c}_{l}^{\dagger}\cop_{l'} + \mbox{H.c.} \right) + V\sum_{\langle l,l' \rangle} \hat{n}_{l}\hat{n}_{l'}
\end{equation}
where the sum over ${\langle l,l' \rangle}$ denotes the set of all nearest neighboring sites, including periodic boundary conditions. $\hat{c}_{l}^{\dagger}$ and $\cop_{l}$ are the fermionic creation and annihilation operators at site $l$ and $\hat{n}_{l} = \hat{c}_{l}^{\dagger}\cop_{l}$ is the local number of fermions at site $l$. We point out the presence of a complex tunneling phase $\phi$, which captures the effect of an artificial gauge field on the system. This phase allows the driving of a non-zero particle current. 

The system is exposed to a dissipative environment, and we describe the system's density operator $\rhop(t)$ to be governed by the following Lindblad master equation \cite{Lindblad, Gorini}
\begin{equation}\label{eq:QME}
\dot{\rhop} = \mathcal{L}(\rhop) =  -\frac{\im}{\hbar}[\Hop, \hat{\rho}] + \Di(\rhop).
\end{equation}
The second term in Eq.(\ref{eq:QME}) describes the evolution due to a dephasing dissipator
\begin{equation}\label{eq:dissipator}
\mathcal{D}(\rhop) = \gamma \sum^{L}_{l} \left(2\;\hat{n}_{l} \rhop \hat{n}_{l}  - \lbrace \nop_{l}^{2}, \rhop \rbrace\right).
\end{equation}
This dissipator has already been used in \cite{BissbortPoletti2017, Pisa} to effectively model friction on a moving particle. This dephasing dissipator is commonly used to model the effects of spontaneous emissions in single band models \cite{Pichler2010, Gerbier2010}. 

As described in the introduction, given the above model, we are interested in the following scenario: Starting from the ground state of the Hamiltonian $\hat{H}(\phi \neq 0)$, we let the system evolve under Eq.(\ref{eq:QME}) with $\hat{H}(\phi = 0)$, i.e. a sudden quantum quench, and study the dynamics of the average particle current $\cura$. This is obtained by computing $\cura (t)=\tr\!\left(\rhop(t)\cur\right)$ where the average current operator \cite{averageJ} is 
\begin{equation}\label{eq:current_operator}
\cur = - \im \frac{J}{\hbar L}\sum^{L}_{\langle l,l'\rangle} \left( \hat{c}_{l}^{\dagger}\cop_{l'} - \hat{c}_{l'}^{\dagger}\cop_{l} \right).
\end{equation}
The average current operator is given by the average of all the bonds' current operators where each bond operator is derived from the continuity equation of the local particle number $d\nop_l/dt=\im/\hbar[\Hop,\nop_l]$.   

In our simulations, given the initial state $\ket{\psi_0}$ and the associated density matrix $\ket{\psi_0}\bra{\psi_0}$, we determine the accessible density matrix subspace within that symmetry sector. Subsequently, we numerically propagate the equations of motion within that subspace and evaluate the observables of interest at various times, each of which can be expressed as a linear form on the vector space of density matrices.

\section{Current decay in presence of interactions}\label{sec:deca}      

The quench from the Hamiltonian $\Hop(\phi)$ to $\Hop(0)$ can induce a current in the system which will then be suppressed by the dissipator. In fact it is easy to see that the unique steady state of this system is the infinite temperature state (or completely mixed state) which has obviously no current. 

The time-dependence of the current can be readily solved analytically in the non-interacting case, $V=0$. To do so we use the Heisenberg picture, which allows us to write the equation of motion for the average current operator as    
\begin{equation}\label{heiseberg_equation}
\frac{d}{dt} \cur = \frac{\im}{\hbar}[\Hop, \cur] + \gamma \sum^{L}_{l} \left(2\;\hat{n}_{j} \cur \hat{n}_{j}  - \lbrace \hat{n}_{j}^{2}, \cur \rbrace\right).
\end{equation}
Since the non-interacting Hamiltonian commutes with the average current operator, this equation can be shown to reduce to
\begin{equation}\label{eq:current_decay}
\frac{d}{dt} \cur = -2 \gamma \cur
\end{equation}
which implies that the average current simply decays exponentially with a decay rate $2 \gamma$. A similar analysis for the kinetic energy was discussed in \cite{BernierKollath2013}.      

In presence of interactions, $\cur$ does not commute with the Hamiltonian and a much richer dynamics is expected.

\subsection{Dependence on the filling}

\begin{figure}
\includegraphics[width=\columnwidth]{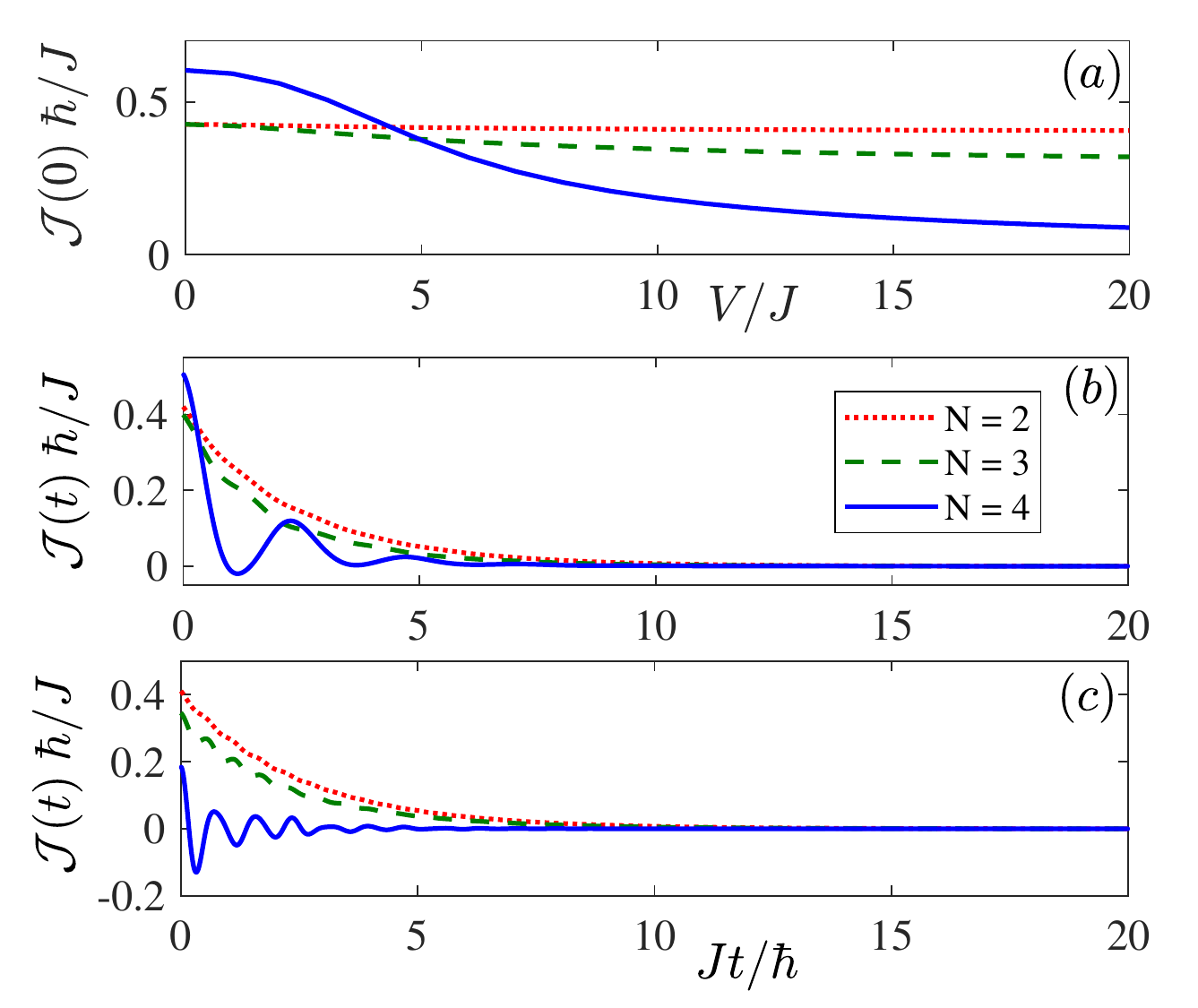}
\caption{(a) Initial average particle current versus interaction strength. (b-c) Time evolution of the average particle current $\mathcal{J}$ after a quench to $\phi = 0$ for interaction strengths (b) $V = 3J$ and (c) $V = 10J$. Common parameters are $\hbar \gamma = 0.2 J$, $\phi = \pi/3$ while the particle number is $N=2$ (red-dotted line), $N=3$ (green-dashed line) or $N=4$ (blue continuous line). } \label{fig:Fig1}   
\end{figure}

We first analyze the dependence of the current on the filling. We concentrate on a chain of $L=8$ sites, which, while being relatively small, allows for the study of different fillings. As naturally expected, the initial current is strongly dependent on the filling and on the interaction. This is analyzed in Fig.\ref{fig:Fig1}(a), where we plot initial current versus the interaction $V$ for different particle numbers in the system. In particular, we consider a total number $N=2$ (red-dotted line), $N=3$ (green-dashed line) and $N=4$ (half-filling, blue continuous line). For low fillings, the initial current is relatively insensitive to the strength of the interaction $V$, while at half-filling, the initial current can be significantly suppressed at large $V$. This is due to the fact that at half-filling all off-site correlations $\langle\cdop_l\cop_{l'}\rangle$ for $l\neq l'$ are suppressed. 
In Fig.\ref{fig:Fig1}(b,c) we plot the average current versus time for different particle numbers. In panel (b) we show the current for $V=3J$, while in panel (c) $V=10J$. In both cases we observe that interactions change the decay of the current from a simple exponential decay (as shown analytically for $V=0$) to a decay with oscillations. These oscillations are more pronounced close to half-filling rather than at low fillings. Moreover, for larger interactions the oscillations are faster.      
  
\subsection{Role of the dephasing rate}

\begin{figure}
\includegraphics[width=\columnwidth]{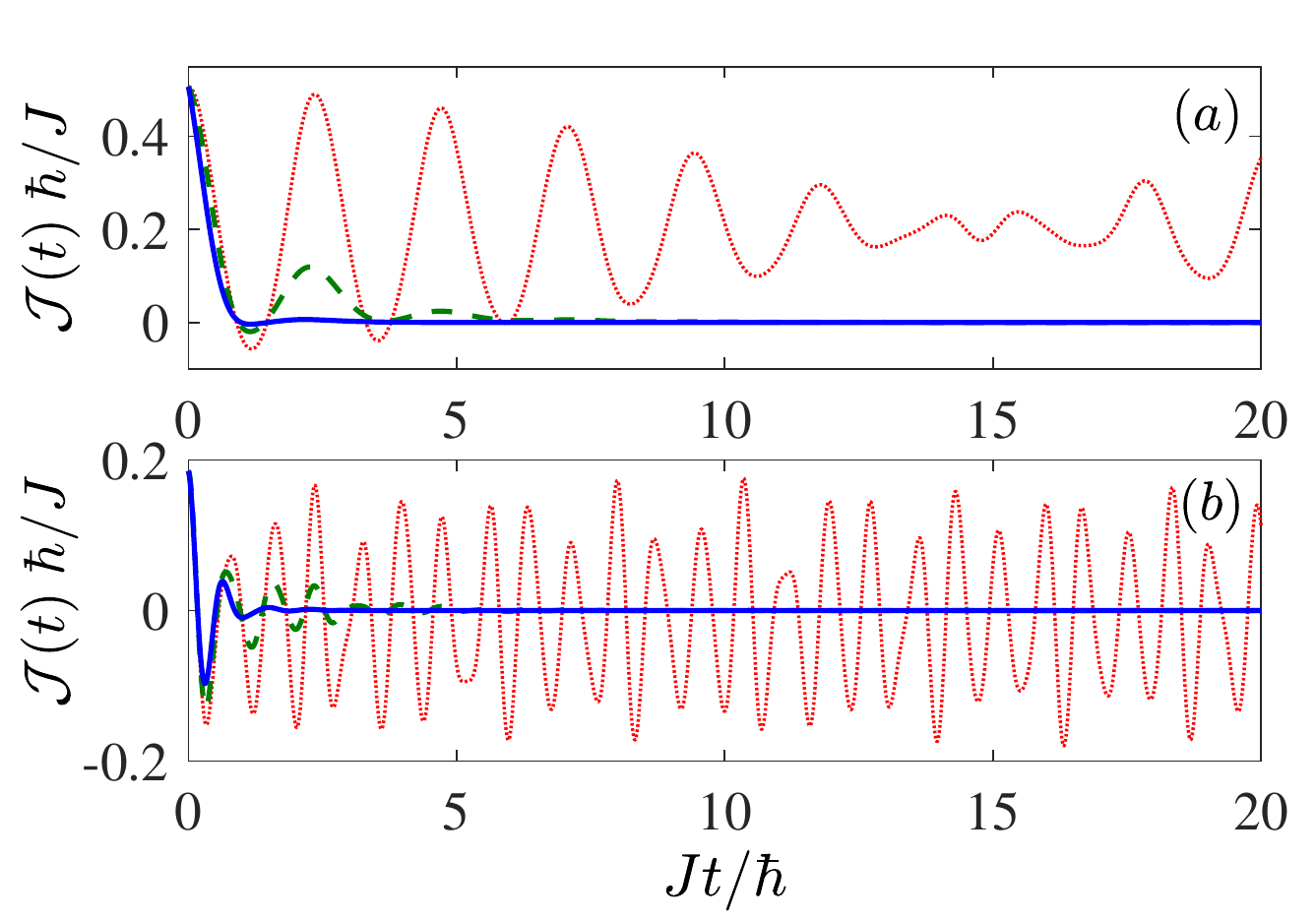}
\caption{Time evolution of the average particle current $\cura$ after a quench from $\phi=\pi/3$ to $\phi = 0$ for various dephasing strengths $\gamma$ and interaction strengths (a) $V = 3J$ and (b) $V = 10J$. Other parameters are $N=4$, $L=8$ and the dephasing rates are $\hbar\gamma=0$ (red dotted lines), $0.2J$ (green dashed lines) and $0.6J$ (blue continuous lines).} \label{fig:Fig3}  
\end{figure} 

The dephasing rate $\gamma$ plays an important role in the decay of the particle current. We analyze this in more detail in Fig.\ref{fig:Fig3}(a,b) for $N=4$, $L=8$ and for $V=3J$, panel (a), and $V=10J$, panel (b). We consider three values of the dissipative rate $\hbar\gamma/J$: $0$ (red dotted lines), $0.2$ (green dashed lines) and $0.6$ (blue continuous lines). For zero dissipation the current does not decay, however, because of the presence of the interaction, it oscillates. For smaller interaction the oscillations are slower and also averaging the current over time gives a non-zero value, indicating that, on average, there is still a net current. For larger interactions the oscillations are faster and much more pronounced, leading to a much smaller time-averaged net current. We also note that if the dissipation is strong enough, it can suppress the oscillations induced by the interaction, and the decay is exponential. This is highlighted in particular by the blue continuous lines in Fig.\ref{fig:Fig3}(a,b) for $\hbar\gamma=0.6J$. It should also be noted that the period of oscillations is slightly modified by the presence of dissipation. This is more clearly visible in Fig.\ref{fig:Fig3}(b) where the curve for $\hbar\gamma/J=0.6$ changes sign at different times compared to the other two curves.

\subsection{Initial current}   

\begin{figure}
\includegraphics[width=0.9\columnwidth]{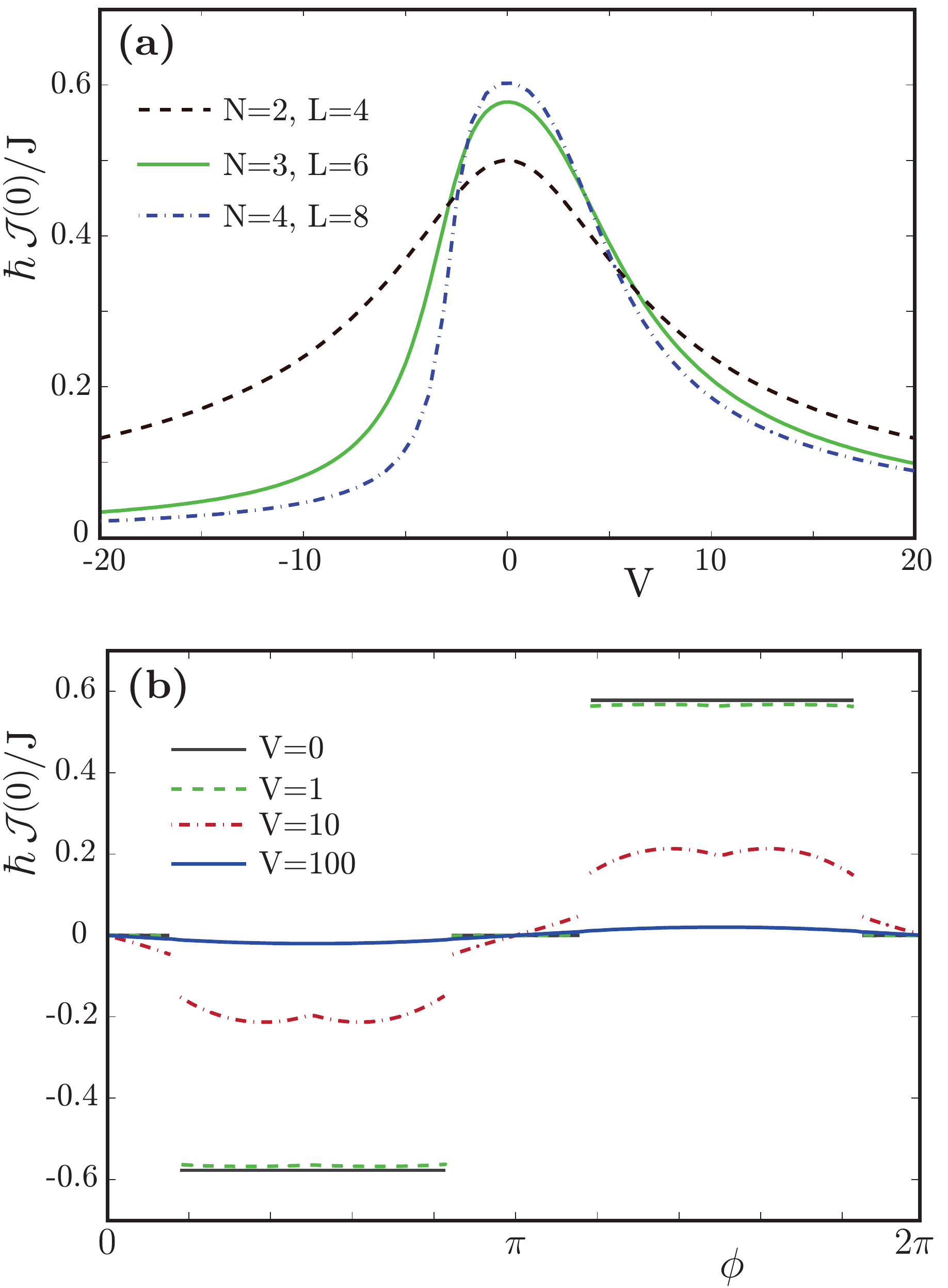}
\caption{(a) Initial current $\mathcal{J}(0)$ as a function of the interaction strength $V$ for various values of the site number $L$ while keeping the particle number as $N=L/2$ and a tunneling phase $\phi=\pi/3$. (b) Initial current $\mathcal{J}(0)$ as a function of the initial phase $\phi$ for various values of the interaction strength $V$ for $N=3$ particles on $L=6$ sites.} \label{fig:init_cur}
\end{figure}

In Fig.\ref{fig:Fig1}(a) we analyze the dependence of the initial current on the strength of the interaction. There we considered a system with a given length and we changed the filling. It is however important to consider the case in which the filling is kept the same while the size of the system increases. We focus on half-filling because, as shown in Fig.\ref{fig:Fig1}(a), the interactions have the strongest effect. Our results are depicted in Fig.\ref{fig:init_cur}(a). While we still observe that the current decreases as the modulus of the interaction increases, we remark a more pronounced asymmetry of the functional form of the initial current $\mathcal{J}(0)$ versus interaction $V$. We also observe that, while at low interaction the current increases with system size $L$, whereas at large interaction the opposite occurs.          

We now consider the dependence of the initial current on the tunneling phase $\phi$, which is shown in Fig.\ref{fig:init_cur}(b) for different values of the interaction strength. This is computed for a system with $L=6$ sites and $N=3$ particles. For $V=0$, it is expected that the ground state suddenly changes for $\phi=\pm\pi/6$, $\phi=\pi\pm\pi/6$ and $\phi=\pi/2$ (this will be discussed in more detail in section \ref{sec:overlap}). This is reflected in large jumps of the current at $\phi=\pm\pi/6$ and $\phi=\pi\pm\pi/6$ and, for $V\neq 0$ a discrete change in the derivative of the current versus interaction. In particular, in Fig.\ref{fig:init_cur}(b) we show the interactions $V=0$ (black continuous line), $V=J$ (green dashed line), $V=10J$ (red dot-dashed line) and $V=100 J$ (blue dotted line). In the limit of very large interactions, the current is strongly suppressed for any value of the initial gauge field $\phi$.  We highlight here that, for $N=2$ and $L=4$, it is possible to show, after much algebra, that the current is symmetric in the interaction strength $V$ \cite{footnote}.

\subsection{Overlap with excited states}\label{sec:overlap}

\begin{figure}
\includegraphics[width=\columnwidth]{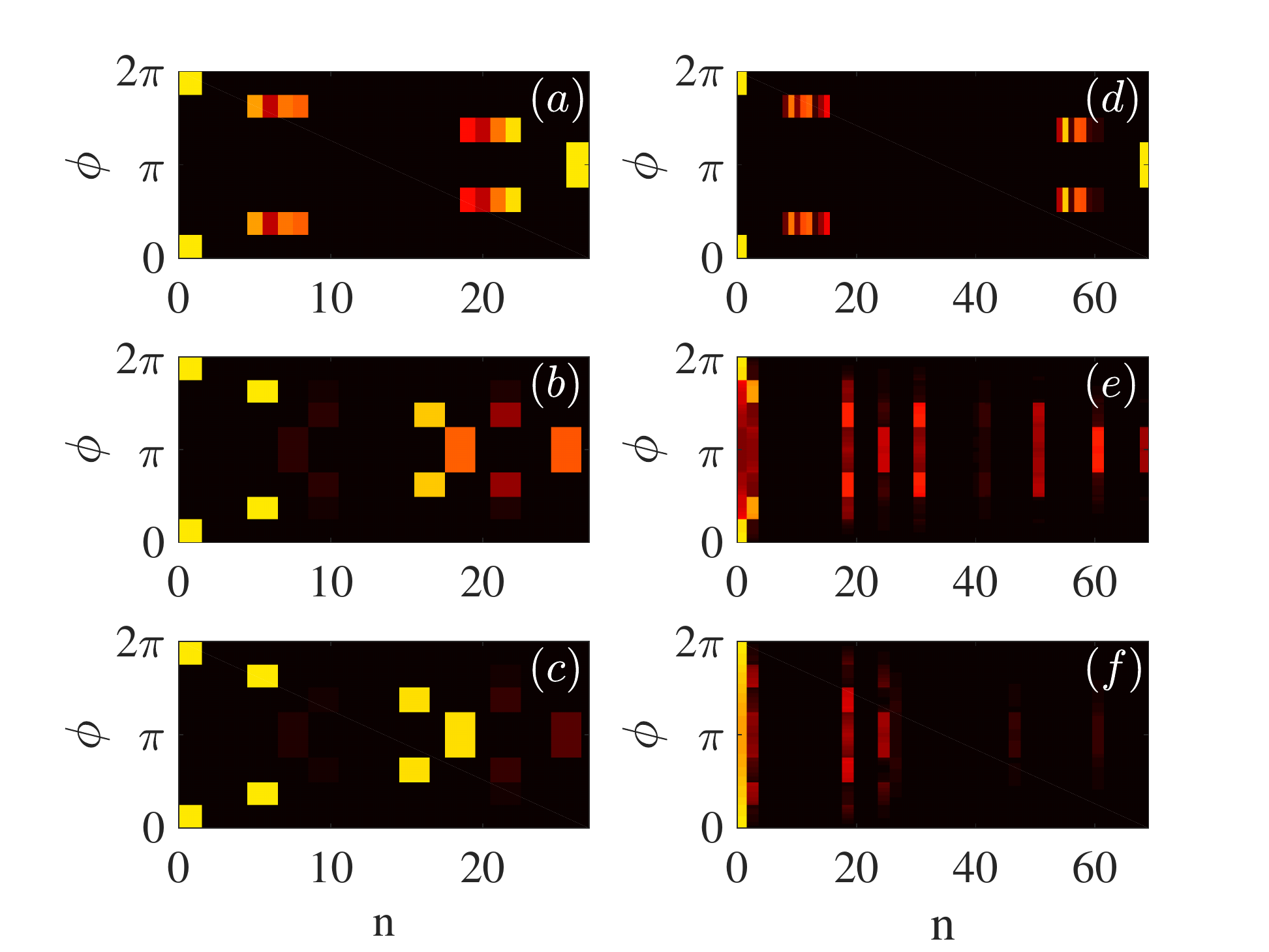}
\caption{Absolute value of the overlaps between the initial state and the eigenstates of the Hamiltonian after the quench, i.e. $|\!\braket{0(\phi)|n(0)}|$ where $\ket{m(\phi)}$ is the $m-$th eigenstates of $\hat{H}(\phi)$ in increasing order of energy. The parameters for the various plots are as follows: (a) $N = 2, V = 0$; (b) $N = 2, V = 3J$; (c) $N = 2, V = 10J$; (d) $N = 4, V = 0$; (e) $N = 4, V = 3J$; (f) $N = 4, V = 10J$.} \label{fig:Fig4}
\end{figure}

The phenomenology just described can be explained by studying the overlap of the initial condition with all the eigenstates of the Hamiltonian for $\phi=0$. 

However, before doing so, it is important to perform a symmetry analysis of the system. For $\phi=2l\pi/L$ where $l$ is an integer number, it is possible to write a unitary transformation which changes $\Hop(0)$ in $\Hop(\phi)=\Uop(\phi)\Hop(0)\Uop^\dagger(\phi)$. This unitary transformation is 
\begin{align}
\Uop(\phi)=\exp\left(\im \sum_m m\phi \nop_m\right)\label{eq:transformation}     
\end{align} 
It is important to highlight that the limitation to $\phi=2l\pi/L$ is due to the system size and boundary conditions. Other values of $\phi$ would give a different phase in the tunneling at the boundary between the last and the first site. 
The presence of such a transformation ensures that the spectrum of $\Hop(2l\pi/L)$ is independent of $l$, while its eigenstates can be computed applying the unitary transformation on the eigenstates of $\Hop(0)$. Moreover, since the Hamiltonian is periodic, $\Hop(\phi)=\Hop(\phi+2\pi)$, it can be derived that the eigenstates belong to different symmetry sectors (similarly to quasi-momentum for a spatially periodic Hamiltonian). For this reason we expect a significant change in the overlap of the ground state of $\Hop(\phi)$ with the eigenstates of $\Hop(0)$.   

In Fig.\ref{fig:Fig4}(a-f) we show an intensity plot of the overlap of the initial condition with the eigenstate of $\Hop(0)$, i.e. $|\langle 0(\phi) | n(0) \rangle|$, where $|m(\phi)\rangle$ represents the $m-$th eigenstate of $\Hop(\phi)$ (in order of energy from the lowest to the largest) \cite{offset}. From this figure we gain insight into the effect of interaction and of filling. In fact, the left column panels (a-c) are computed for a system for $N=2$, while panels (d-f) for the half-filling case $N=4$. Moreover, panels (a,d) are for the non-interacting case $V=0$, (b,e) for an intermediate interaction $V=3J$, while panels (c,f) are for the strong interacting case $V=10J$. In Fig.\ref{fig:Fig4}(a-c) we observe that, as the phase is varied, the overlap of the initial state with the various eigenstates shifts abruptly, as structurally different states compete to be the ground state. This dependence of the overlap on $\phi$ leads to the observed oscillations in the time-evolution of the current. In Fig.\ref{fig:Fig4}(d-f), for $N=4$, e.g. at half-filling, the change in overlap between the various panels is even more marked. In particular, comparing Fig.\ref{fig:Fig4}(d) and (f) we notice a significant difference, due to the strong role of interactions at half-filling: While for $N=2$, panel (d), the overlap changes significantly with $\phi$, at half-filling ($N=4$), panel (f), the overlap is mostly concentrated on the ground state. Since the ground state has zero current, this explains why the current is significantly reduced in the strongly interacting regime at half-filling.

\subsection{Frequency of current oscillations}

\begin{figure*}
\includegraphics[width=\linewidth]{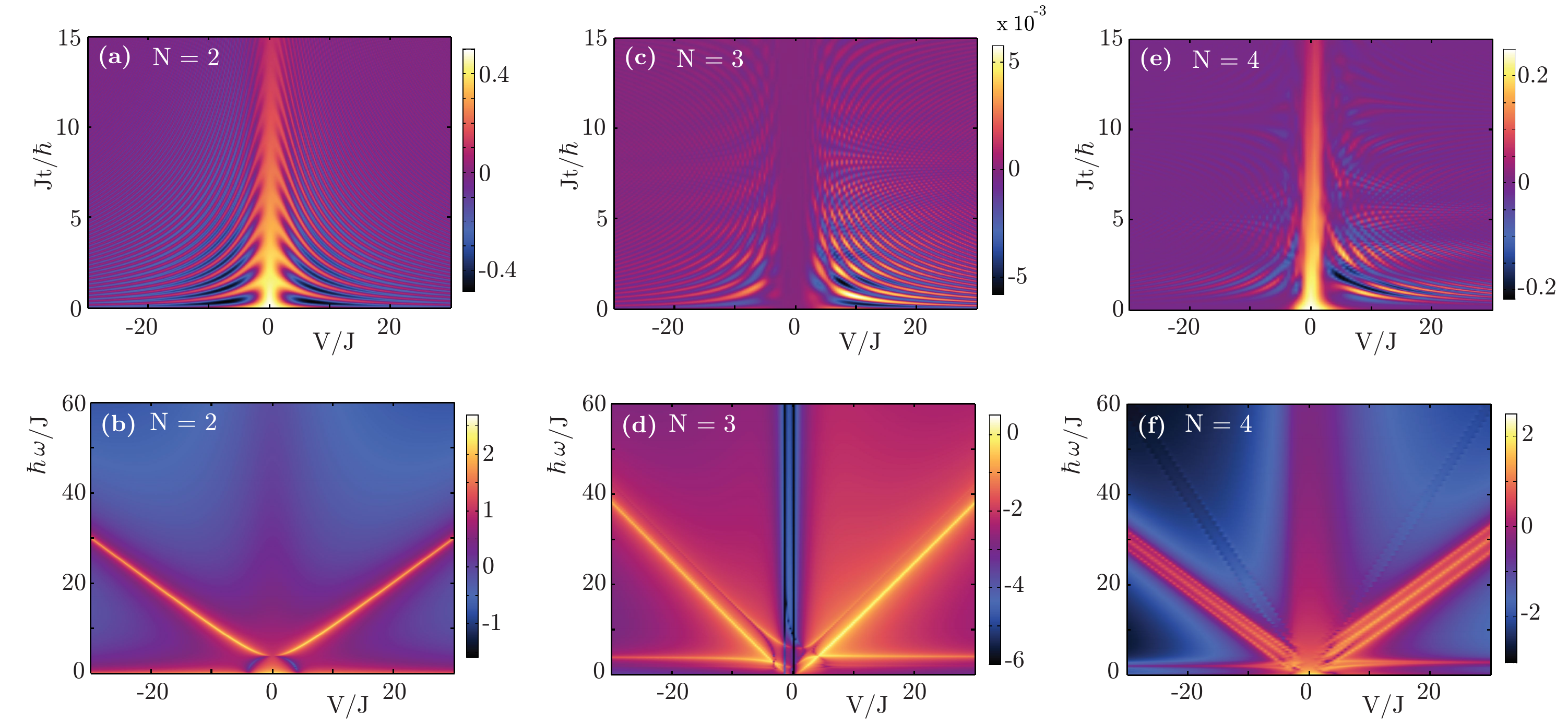} \caption{(a,c,e) Dependence of time-dependent current $\mathcal{J}(t)$ (in units of $\hbar/J$) on the interaction strength $V$ at half filling, $L=2N$. (b,d,f) The associated Fourier transforms $\log_{10}|\mbox{FT}[\hbar\;\mathcal{J}(t)/J]|$ (shown on a logarithmic color scale for clarity) reveal the relevant frequencies in the dynamics and their dependence on the interaction strength $V$. Common parameters are $\hbar\gamma=0.05J$ and $\phi=3.1$, while $L=4$ (a,b), $L=6$ (c,d) and $L=4$ (e,f).} \label{fig:Fig5}
\end{figure*}

We now discuss in more detail the temporal evolution of the current as a function of the interaction for a given initial tunneling phase $\phi=3.1$ (this value has been chosen to avoid degeneracies). In Fig.\ref{fig:Fig5}(a,c,e) we show $\mathcal{J}(t)$ versus interaction strength $V$ at half-filling ($N=L/2$) for different system sizes: (a) $L=4$, (c) $L=6$ and (e) $L=8$. These plots clearly show that for $V=0$ there are no oscillations and these become more pronounced with a frequency that increases with the interaction strength. 

We also study the Fourier transform of $\mathcal{J}(t)$ to gain information on the relevant frequencies $\omega$. $\mathcal{J}(\omega)$ is depicted in Fig.\ref{fig:Fig5}(b,d,f) respectively for Fig.\ref{fig:Fig5}(a,c,e). In each figure we observe at least a peak which is proportional to the interaction, although parallel lines and higher order peaks can also be observed (most notably in Fig.\ref{fig:Fig5}(b,d,f). 
In fact, above the ground state there are excitations with an energy given approximately by $V$ (and multiples), but which between them are separated by an energy of the order $J$ due to the tunneling term of the Hamiltonian. This explains the presence of various parallel peaks and also the presence of a peak at low frequencies which is independent of $V$. This peak is due to the beating of the modes with approximate energy $V$.           
We remark that the color intensity for Fig.\ref{fig:Fig5}(d-f) is shown on a logarithmic scale ($\log_{10}(|\mbox{FT}[\hbar\;\mathcal J(t)/J]|)(\omega)$) to better show the details of the features of the current. However, in this way it is harder to notice the suppression of the overall magnitude of the current.    

At small $V/J$ there is an emergence of multiple non-trivial oscillation frequencies which cannot be attributed to either $V$, $J$ or $\gamma$ individually (see, for instance, Fig.\ref{fig:Fig5}(d) near $V=0$).

\section{Conclusions}\label{sec:conclusions}

This study gives insight on the response of an interacting quantum system to a current-inducing quench and to a current-dissipating environment. We show that interactions can induce oscillations in the decay of the current which are faster for larger interactions. The functional form of the decay of the current is, in general, significantly affected by the gauge field of the initial preparation because of the different overlap of the initial state with the eigenstates of the Hamiltonian. The effect of interactions is more pronounced when the system is close to half-filling, as the spectrum and the eigenstates of the Hamiltonian are significantly affected by interactions \cite{qpt}. Specifically, at half-filling the overlap of the initial state with the ground state of the Hamiltonian after the quench is largest and mostly insensitive to the tunneling phase $\phi$. We have also shown that the dephasing rate $\gamma$, in addition to induce a decay, also leads to a renormalization of the oscillations frequency.   

We find a strong dependence of the system dynamics on the system size and filling factor. Since we are interested in fermions in small periodic optical lattices, in the current work the finite size is seen as a feature. However it would also be relevant to study a similar set-up in the continuum and for larger dimensional systems. Further studies could focus on the use of different baths, for instance finite temperature baths, or baths which could help to sustain or even induce a current. Along this line, of particular importance is the interplay between the symmetries of the Hamiltonian and of the dissipator \cite{ManzanoHurtado2017}.

{\it Acknowledgments:} D.P. acknowledges fruitful discussions with D. Ho and L. Santos. This work was supported by the Air Force Office of Scientific Research under Award No. FA2386-16-1-4041 and Singapore MOE Academic Research Fund Tier-2 project (Project No. MOE2016-T2-1-065).


\begin{thebibliography}{100}
\bibitem{averageJ} We have defined here the average current operator because it is the convenient operator to use for deriving Eq.(\ref{eq:current_decay}). However, given the initial conditions and the Hamiltonians used in this work, the current is identical at each bond.
\bibitem{footnote} We write the $6-$dimensional Hamiltonian in a gauge in which the tunneling phase is non-zero only in the coupling between the first and the fourth site, and, for instance, we compute the current for one of the eigenstates of the preparation Hamiltonian. This results in  
$$\mathcal{J}(0)= -\frac{8J}{\hbar} \frac{ f(V/J,\phi) \left( \sin(2\phi)^2 + \vert\sin(2\phi)\vert  \right)/\sin(4\phi) }{ 16 + f(V/J,\phi)^2 (1+\vert\sin(2\phi)\vert)/\cos(2\phi)^2}$$ 
with 
$f(x,\phi)=x-\sqrt{16+x^2 -16\vert\sin(2\phi)\vert}$. This expression of the current is symmetric in $V$.
\bibitem{offset} We point out that in this system the ground state can be degenerate for certain values of $\phi$. For this reason we choose initial gauge fields which are off-set from these degenerate points.
\bibitem{qpt} Note that at half-filling there is a quantum phase transition in the limit of $L\rightarrow \infty$.
\end{thebibliography}
\end{document}